\documentclass[twocolumn, switch]{article} 
\usepackage{preprint}
\usepackage{comment}
\usepackage{amsmath, amsthm, amssymb, amsfonts}

\usepackage[numbers,square]{natbib}

\usepackage[utf8]{inputenc}	
\usepackage[T1]{fontenc}	
\usepackage{xcolor}		
\usepackage[colorlinks = true,
            linkcolor = purple,
            urlcolor  = blue,
            citecolor = cyan,
            anchorcolor = black]{hyperref}	
\usepackage{booktabs} 		
\usepackage{nicefrac}		
\usepackage{microtype}		
\usepackage{lineno}		
\usepackage{float}			

\usepackage{lipsum}		

\usepackage{newfloat}
\DeclareFloatingEnvironment[name={Supplementary Figure}]{suppfigure}
\usepackage{sidecap}
\sidecaptionvpos{figure}{c}

\usepackage{titlesec}
\titlespacing\section{0pt}{10pt}{1pt}
\titlespacing\subsection{0pt}{10pt plus 3pt minus 3pt}{1pt plus 1pt minus 1pt}
\titlespacing\subsubsection{0pt}{8pt plus 3pt minus 3pt}{1pt plus 1pt minus 1pt}

\usepackage{tikz,xcolor,hyperref}

\definecolor{lime}{HTML}{A6CE39}

\title{Bounds reach of a future circular collider on the parameters of a minimal Baryon-Lepton symmetric model}

\usepackage{background}
\usepackage{xcolor}     

\backgroundsetup{
  scale=1,
  color=gray!90,
  opacity=1,
  angle=0,
  position=current page.south,
  vshift=1.2cm,
  contents={\texttt{*correspondence: marlon.brade@g.msuiit.edu.ph}}
}

\usepackage{authblk}

\author{Marlon P. Brade$^{1,2,3}$}
\author{Jeremiah D. Juevesano$^{1,2,4}$}
\author{Dennis C. Arogancia$^{1,2}$}
\author{Jan Mickelle V. Maratas$^{1,2}$}

\affil{$^{1}$MSU-Iligan Institute of Technology, Premier Research Institute of Science and Mathematics (PRISM), Iligan City, Philippines}

\affil{$^{2}$MSU-Iligan Institute of Technology, Department of Physics, Iligan City, Philippines}

\affil{$^{3}$Caraga State University, Department of Physics, Ampayon, Butuan City, Philippines}

\affil{$^{4}$University of the Philippines - Mindanao, Department of Mathematics, Physics, and Computer Science, Davao City, Philippines}

\makeatletter
\begin{document}
\twocolumn[ 
  \begin{@twocolumnfalse}
  
\maketitle

\begin{abstract}
This paper investigates the physics potential of the proton–proton Future Circular Collider (FCC-hh) in the search for a new heavy gauge boson, \( Z' \), within the framework of the minimal Baryon-Lepton (\( B-L \)) symmetric model. We examine the exclusion limits on the mass of the \( Z' \) boson for various sets of coupling constants, utilizing leptonic decays, specifically \( \ell^+\ell^- \). Using the \texttt{DARKCAST} framework, we compare the parameter constraints derived from previous experiments with the expected reach of the FCC-hh, providing insights into the collider's sensitivity and potential to probe new physics beyond the Standard Model.
\end{abstract}
\vspace{0.35cm}

  \end{@twocolumnfalse} 
] 



\section{Introduction}

The Standard Model (SM) of particle physics, though remarkably successful in describing the fundamental particles and their interactions, leaves several key phenomena unexplained. It does not account for the observed dark matter abundance, the origin of neutrino masses and oscillations, the hierarchy problem, or the matter–antimatter asymmetry of the Universe \cite{Hernandez:2025spl,Hindmarsh:2020hop}. In addition, persistent tensions in lepton-flavor universality measurements \cite{Seuthe:2022vea} provide further motivation to search for physics beyond the Standard Model (BSM).

In recent years, considerable attention has been devoted to extensions of the SM that introduce an additional $U(1)$ gauge symmetry. Such scenarios naturally give rise to a new neutral gauge boson, commonly denoted as $Z'$, and have been widely studied as simple yet well-motivated frameworks for BSM. Among these, one of the most economical and theoretically appealing choices is the baryon minus lepton number ($B\!-\!L$) extension. In order to cancel gauge anomalies, this model predicts the existence of three additional leptons, identified as right-handed neutrinos, which are absent in the SM. Along with these neutrinos, the model introduces an additional Higgs singlet $\chi$ and the new gauge boson $Z'$. The singlet $\chi$ carries $B\!-\!L$ charge and couples to the right-handed neutrinos via Yukawa interactions, thereby generating their masses through the Higgs mechanism. An interesting feature of the $B\!-\!L$ model is its ability to yield both positive and negative contributions to the anomalous magnetic moment of leptons by tuning the vector couplings of fermions to the $Z'$ \cite{Bodas:2021fsy}. Furthermore, the $B\!-\!L$ framework provides a natural setting for baryogenesis through leptogenesis \cite{Abbas:2007ag,Iso:2010mv}, offering a compelling explanation for the observed matter–antimatter asymmetry in the universe.

A broad experimental program has already placed important constraints on new neutral gauge bosons (\(Z'\)) arising from extended \(U(1)\) symmetries, including the minimal Baryon minus Lepton (\(B\!-\!L\)) scenario. Leptonic colliders such as LEP carried out extensive searches for new particles, including possible signatures of a \(Z'\), but no signal has been observed. These precision measurements translate into indirect exclusion limits on the \(Z'\) mass and coupling, typically requiring \(M_{Z'}/g' \gtrsim 7~\text{TeV}\)~\cite{ALEPH:2013dgf,Cacciapaglia:2006,Das:2021esm}. Beam-dump, fixed-target, and low-energy scattering experiments have probed the MeV–GeV mass range for signs of new gauge interactions, but no signal was detected, leading to exclusion limits in that region~\cite{A1:2011yso,NA64:2019auh,NA64:2017vtt,KLOE-2:2016ydq,APEX:2011dww,Asai:2023mzl,Essig:2010xa,Freytsis:2009bh}. Collider experiments such as BaBar and the LHC have provided the most stringent bounds on the \(Z'\) mass in the multi-GeV to multi-TeV range, further restricting the available parameter space~\cite{BaBar:2014zli,BESIII:2017fwv,LHCb:2017trq,CMS:2020ulv,CMS:2024zqs}. Additionally, Constraints have also been obtained from the cosmic frontier, where observations of the cosmic microwave background, high-energy neutrino measurements from IceCube, Lyman-$\alpha$ data, and other cosmological experiments impose additional limits on the parameter space of the $Z'$ \cite{KA:2023dyz, Barman:2024lxy, Barman:2025bir,Barman:2025hoz}.

Despite decades of experimental effort, the direct discovery of dark matter, new gauge bosons such as the \(Z'\), or any particle of BSM has remained elusive. The Large Hadron Collider (LHC) has played a central role in testing the Standard Model and searching for new physics at the TeV scale. While it has not provided evidence for new particles beyond the Higgs boson, the LHC has established stringent limits on many extensions of the SM, including searches for additional neutral gauge bosons. To further extend the energy frontier and to explore parameter regions that remain inaccessible, future collider projects have been proposed. Among these, the hadronic Future Circular Collider (FCC-hh) is designed to reach a center-of-mass energy of about 100~TeV with an integrated luminosity of 30~ab\(^{-1}\). Such capabilities would allow precision measurements of SM processes at unprecedented scales and dramatically improve the discovery potential for heavy new particles such as the \(Z'\). In parallel, the proposed International Linear Collider (ILC) is designed to operate as a high-precision \(e^+e^-\) facility with center-of-mass energies up to \(1~\text{TeV}\). In the context of the \(B\!-\!L\) model, these precision measurements are particularly relevant, as the anomaly-free realization of the gauge symmetry implies an upper theoretical limit on the parameter ratio \(M_{Z'}/g' \gtrsim 137.2~\text{TeV}\) \cite{Das:2021esm,Babich:2010zz, Pankov:2017dkv}. The ILC therefore provides a complementary probe of the model by constraining the coupling structure even when the \(Z'\) mass lies well above its direct kinematic reach. Other proposed facilities, including the High-Luminosity LHC (HL-LHC)~\cite{Apollinari:2015wtw,LHeC:2020van}, the High-Energy LHC (HE-LHC)~\cite{FCC:2018bvk}, and the Circular Electron–Positron Collider (CEPC)~\cite{CEPC-SPPCStudyGroup:2015csa,CEPCStudyGroup:2023quu}, provide important complementary opportunities. Nevertheless, among these, the FCC-hh offers the broadest reach for probing uncharted regions of parameter space in both the SM and its extensions~\cite{FCC:2018evy,FCC:2018vvp,FCC:2018byv,Armesto:2014iaa}.

The following sections are organized as follows. In the next chapter, we provide a detailed discussion of the $B\!-\!L$ model, outlining its theoretical structure and phenomenological features relevant for collider studies. The subsequent chapter then presents the simulation setup adopted in this work, including the choice of parameters, event generation, and analysis strategy used to evaluate the sensitivity of the FCC-hh to the $Z'$.

\section{Minimal Baryon - Lepton Model}

The B-L model is constructed upon the framework of gauge symmetry, $SU(3)_C \otimes SU(2)_L \otimes U(1)_{Y} \otimes U(1)_{B-L}$ and is considered a triply-minimal extension of the SM. Firstly, it retains a level of minimality within the gauge sector by introducing a $U(1)_{B-L}$ symmetry, which preserves the $B-L$ number. Furthermore, in the fermionic sector, the model exhibits minimalism by introducing a lone singlet fermion for each generation to address the $U(1)_{B-L}$ gauge anomalies. These fermions can be naturally interpreted as the right-handed neutrinos. Finally, within the scalar sector, it maintains minimalism as it introduces a neutral scalar singlet that plays a pivotal role in spontaneously breaking the $U(1)_{B-L}$ symmetry responsible for the mass generation of the additional $Z'$ gauge boson. The Lagrangian of the $B-L$ model can be decomposed as follows, 

\begin{equation}
    \mathcal{L}_{B-L} =  \mathcal{L}_{S} + \mathcal{L}_{YM} + \mathcal{L}_{f} + \mathcal{L}_{Y},
\end{equation} 

\noindent where $\mathcal{L}_{S}$ is the scalar Lagrangian,  $\mathcal{L}_{YM}$ is the Yang-Mills Lagrangian,  $\mathcal{L}_{f}$ is the fermionic Lagrangian and  $\mathcal{L}_{Y}$ is the Yukawa Lagrangian. Brief discussion about Yang-Mills Lagrangian, scalar Lagrangian and spontaneous breaking of B-L symmetry is provided below but detailed discussion can be found in Refs. \cite{Basso:2010pe,Basso:2008iv,Basso:2010jm}.

\subsection{Yang-Mills Lagrangian}
\label{sec:2}

The Yang-Mills Lagrangian of the B-L model is written as follows,

\begin{equation}
    \mathcal{L}_{YM} = -\frac{1}{4} F^{\mu\nu} F_{\mu\nu} - \frac{1}{4} F'^{\mu\nu} F'_{\mu\nu}
\end{equation}

\noindent where,

\begin{center}
   $F_{\mu\nu} = {\partial_\mu{B_\nu}} - {\partial_\nu{B_\mu}}$ \\
   $F'_{\mu\nu} = {\partial_\mu{B'_\nu}} - {\partial_\nu{B'_\mu}}$. 
\end{center}

\noindent The fields $B_\mu$ and $B'_\mu$ are the gauge fields of $U(1)_Y$ and $U(1)_{B-L}$. In this field basis, the covariant derivative is defined as,

\begin{equation}
\begin{split}
    D_\mu \equiv \partial_\mu + ig_s T^\alpha G^\alpha_\mu +  ig T^a W^a_\mu  + i g_1 Y B_\mu  \\
     + i(\tilde{g} Y + g'Y_{B-L})B'_\mu    
\end{split}
\end{equation}

\noindent where $g_s$ is the SU(3)$_C$ gauge coupling, with $T^\alpha$ and $G^\alpha_\mu$ denoting the SU(3) generators and gluon fields, respectively. The term $g$ is the SU(2)$_L$ coupling with $T^a$ as its generators and $W^a_\mu$ as the weak bosons. The U(1)$_Y$ hypercharge interaction involves the coupling $g_1$, the hypercharge $Y$, and the gauge boson $B_\mu$. The final term represents the extended gauge interaction with an extra U(1)$'$ symmetry, where $g'$ is the new coupling constant, $Y_{B-L}$ is the $B-L$ charge, $B'_\mu$ is the new gauge boson, and $\tilde{g}$ is the kinetic mixing parameter that introduces mixing between the hypercharge and U(1)$'$ fields. The minimal $B-L$ model is defined for the case where $\tilde{g} = 0$, which ensures no mixing between SM $Z^0$ boson and the B-L  $Z'$ boson at tree level.

\subsection{Spontaneous Symmetry Breaking }
\label{sec:4}

By breaking the electroweak (EW) and $U(1)_{B-L}$ symmetries and diagonalizing the gauge boson mass matrix, we can arrive to the basis of mass eigenstates, $A_\mu$, $Z_\mu$ and $Z'_\mu$, expressed as follows,
\begin{equation}   
{\begin{pmatrix} B^\mu  \\ W_3^\mu  \\ B'^\mu \end{pmatrix} }  =  \begin{pmatrix} c_{\theta}  & -s_{\theta}c_{\theta'} & s_{\theta} s_{\theta'} \\ s_{\theta}  & c_{\theta}c_{\theta'} & -c_{\theta}s_{\theta'} \\ 0 & s_{\theta'} & c_{\theta'}  \end{pmatrix} {\begin{pmatrix} A^\mu  \\  Z^\mu  \\ Z'^\mu \end{pmatrix}, }
\end{equation}
\noindent with $\frac{-\pi}{4} \leq \theta' \leq \frac{\pi}{4} $, such that:

\begin{equation}
    \tan{2\theta'} = \frac{2\tilde{g} \sqrt{g^2 + g_1^2}}{\tilde{g}^2 + 16(\frac{x}{v})^2g'^2 - g^2 - g_1^2}.
\end{equation}

\noindent In the above expressions, note that $c(s)$ denotes cosine (sine), $\theta$ is the weak-mixing angle, $\theta'$ is the $Z^0-Z'$ mixing angle and the parameter \( v \) denotes the vacuum expectation value (vev) of the Standard Model (SM) Higgs doublet, which spontaneously breaks the electroweak symmetry SU(2)$_L \times$ U(1)$_Y$ down to U(1)$_\text{em}$. The parameter \( x \), on the other hand, is the vev of a new scalar field responsible for breaking the additional U(1)$'$, associated with the gauged \( B-L \) symmetry. By imposing that no mixing between SM $Z^0$ boson and the B-L  $Z'$ boson at tree level, then $\tilde{g} = 0$. The simplified expressions for the masses of gauge bosons are as follows, 

\begin{align}
     M_A &= 0 \\
   M_Z &= \sqrt{g^2 + g_1^2} \cdot \frac{\nu}{2}  \\
   M_{Z'} &= 2g'x.
\end{align}

\section{Simulation Setup}

\subsection{Monte Carlo Generation}
\label{sec:6}
The search of $Z'$ is done in the  leptonic ($\ell^+ \ell^-$) decay channel. \textsf{Monte Carlo} (MC) signal and background events are simulated using \textsf{MG5\_aMC} \cite{Alwall:2014hca} version 3.1.0, with the \textsf{NNPDF23\_lo\_as\_0130\_qed} PDF set \cite{NNPDF:2014otw}.

\subsection{Detector Response}

The detector response has been simulated via the \textsf {DELPHES}\cite{deFavereau:2013fsa} software package where default FCC-hh with pile-up card of \textsf{DELPHES} version 3.5.0 was used. Following a collision event, which involves parton showering, hadronization, and particle decays, \textsf{DELPHES} initiates its process by ensuring the propagation of long-lived particles through the tracking volume. This volume is situated within a uniform axial magnetic field, aligned parallel to the beam direction. The resolution $\sigma$ on the track transverse momentum $p_T$, is described as,


\begin{equation}
    \frac{\sigma(p_T)}{p_T} \approx \frac{\sigma_{r\phi} p_T}{B \cdot L^2},
\end{equation}

\noindent where $B$ is the magnetic field strength, $L$ is the size of the tracking radius and $\sigma_{r\phi}$ is the single hit spatial resolution. Following their propagation within the magnetic field, long-lived particles subsequently arrive at the electromagnetic (\textsf{ECAL}) and hadronic (\textsf{HCAL}) calorimeters where particles deposits their energy. The resolutions $\sigma$ of \textsf{ECAL} and \textsf{HCAL} are parameterized as a function of the particle energy and the pseudorapidity given by the expression,

\begin{equation}
    \left( \frac{\sigma}{E} \right)^2 = \left( \frac{S(\eta)}{\sqrt{E}} \right)^2 + \left( \frac{N(\eta)}{E} \right)^2 + C(\eta)^2,
\end{equation}

\noindent where $S$, $N$ and $C$ are respectively the stochastic, noise and constant terms. Subsequently, the next step in \textsf{DELPHES} is the reconstruction of particles as objects. The efficiencies for triggering, reconstructing, and identifying these particles are parameterized as functions of their momentum where a universal parameterisations  is employed for each object to simplify the process.

\subsection{Statistical Analysis}
\label{sec:8}

 Hypothesis testing is conducted by employing a modified frequentist method based on profile likelihood \cite{Cowan:2010js,Moneta:2010pm}. This approach takes into consideration the systematic uncertainties as nuisance parameters, which are adjusted to align with the expected Monte Carlo.  The profile log-likelihood ratio test statistic is defined as, $q_\mu = -2ln(\mathcal{L}(\mu,\hat{\hat{\theta}})/ \mathcal{L}(\hat{\mu},\hat{\theta} ))$. Note here that $\hat{\theta}$ and $\hat{\mu}$ are the values of the parameters  that maximized the likelihood function and $\hat{\hat{\theta}}$ are the values of the nuisance parameters that maximize the likelihood function for a given value of $\mu$. When there are no substantial deviations from the background expectation, $q_\mu$ is employed within the $CL_s$ method to establish a 95\% Confidence Level (CL) upper limit on the product of the signal production cross-section times branching ratio. For a given hypothesis whose values of the production cross section times branching ratio parameterized by $\mu$  gives $CL_s < 0.05$, these hypothesis are excluded at 95\% CL.

\subsection{Determination of \((M_{Z'}, g')\) Bounds}
\label{sec:9}

To constrain the model parameters, we compare the theoretical predictions of the \(Z'\) production cross section with the expected experimental sensitivities. The limits on the parameter space \((M_{Z'}, g')\) are determined by evaluating the 95\% confidence level (CL) upper limits on the cross section for each considered \(Z'\) mass using the \(\mathrm{CL_s}\) statistical method. 

For each mass point, the corresponding coupling strength \(g'\) is then obtained by iteratively adjusting its value in the event generation software until the predicted cross section \(\sigma(pp \to Z') \times \mathrm{BR}(Z' \to \ell^+\ell^-)\) equals the expected upper limit derived from the \(\mathrm{CL_s}\) procedure. This iterative approach effectively maps the exclusion contour in the \((M_{Z'}, g')\) plane, providing a direct correspondence between the collider sensitivity and the theoretical model parameters.

\begin{figure}[!h]
\includegraphics[width=0.50 \textwidth]{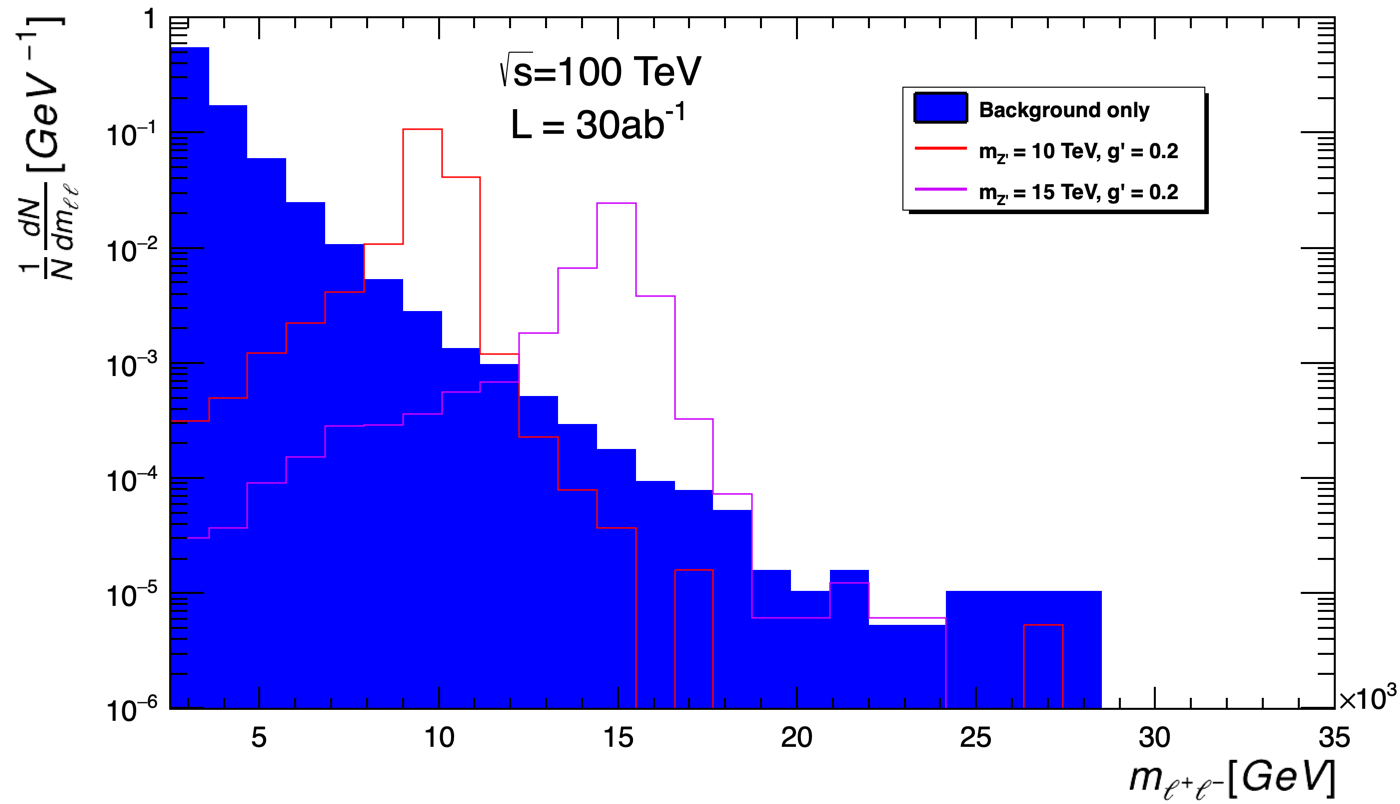}
\caption{Invariant mass  for the case of $M_{Z'} = 10$ TeV and $g' = 0.2 $ in the $\ell^+ \ell^-$ channel.}
\label{fig1}
\end{figure}

\section{Results and discussion}
\label{R&D}

In this analysis, we select the two leptons with the highest transverse momentum (\(p_T\)), ensuring they have opposite electric charges and the same flavor. We then impose the following criteria: each lepton must have \(p_T > 1.0~\text{TeV}\), lie within the pseudorapidity range \(|\eta| < 4.0\), and the dilepton invariant mass must satisfy \(m_{\ell\ell} > 2.5~\text{TeV}\). These event selections are motivated by previous analyses performed at the Future Circular Collider (FCC) that explored heavy neutral gauge bosons in the dilepton channel~\cite{Helsens:2019bfw}. In this study, only the dominant Standard Model Drell--Yan background process (\(pp \rightarrow Z/\gamma^* \rightarrow \ell^+\ell^-\)) is considered, as it constitutes the primary source of dilepton events at hadron colliders. The Drell--Yan cross-section decreases rapidly with increasing invariant mass, resulting in a very small background yield once the \(m_{\ell\ell} > 2.5~\text{TeV}\) requirement is applied. This ensures that the signal region at high mass is effectively background-free, allowing a clear separation between a possible \(Z'\) resonance and the Standard Model expectation.

\begin{figure}[!h]
\includegraphics[width=0.50 \textwidth]{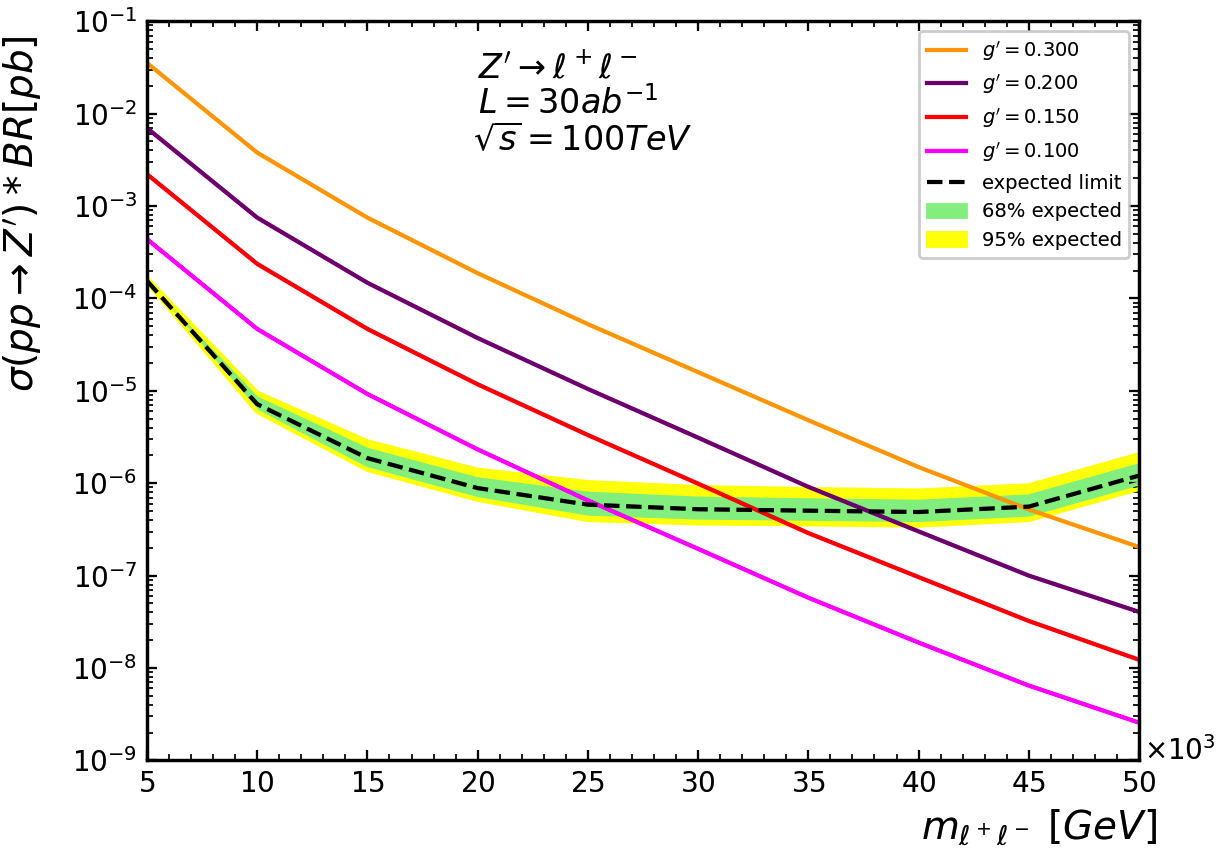}
\caption{95\% CL limit versus mass for the combined dilepton (ee, $\mu\mu$) channel. The colored solid curves correspond to different values of the coupling constant (\(g' = 0.1\)–0.3), while the dashed black line represents the median expected limit derived from the background-only hypothesis. The green and yellow bands indicate the 68\% and 95\% expected confidence intervals, respectively.}
\label{fig2}
\end{figure}

Figure~\ref{fig1} presents the normalized invariant-mass distributions for the combined dilepton (\(e^+e^- + \mu^+\mu^-\)) final state at \(\sqrt{s} = 100~\text{TeV}\) and an integrated luminosity of \(30~\text{ab}^{-1}\). The shaded blue histogram corresponds to the Drell--Yan background, while the overlaid lines indicate the expected signal for benchmark scenarios with \(M_{Z'} = 10~\text{TeV}\) and \(15~\text{TeV}\) assuming \(g' = 0.2\). As shown, the Drell--Yan contribution decreases rapidly with increasing invariant mass, producing an exponentially falling spectrum. In contrast, the \(Z'\) signal appears as a distinct resonance peak at the corresponding mass values, clearly emerging above the smoothly falling background. This demonstrates that the chosen selection criteria are highly effective in isolating potential high-mass resonances from the Standard Model expectation. The resulting statistical sensitivity of the analysis is summarized in Figure~\ref{fig2}, which shows the 95\% CL upper limits on the production cross section times branching ratio, \(\sigma(pp \rightarrow Z') \times \text{BR}(Z' \rightarrow \ell^+\ell^-)\), as a function of the \(Z'\) mass. For the largest coupling considered (\(g' = 0.3\)), the analysis excludes \(Z'\) masses up to approximately \(45~\text{TeV}\) at the 95\% CL. This result highlights the exceptional reach of the FCC in probing heavy neutral gauge bosons through the dilepton channel, where the near absence of background at multi-TeV invariant masses provides a clean environment for discovery.

\begin{figure}[!h]
\includegraphics[width=0.50 \textwidth]{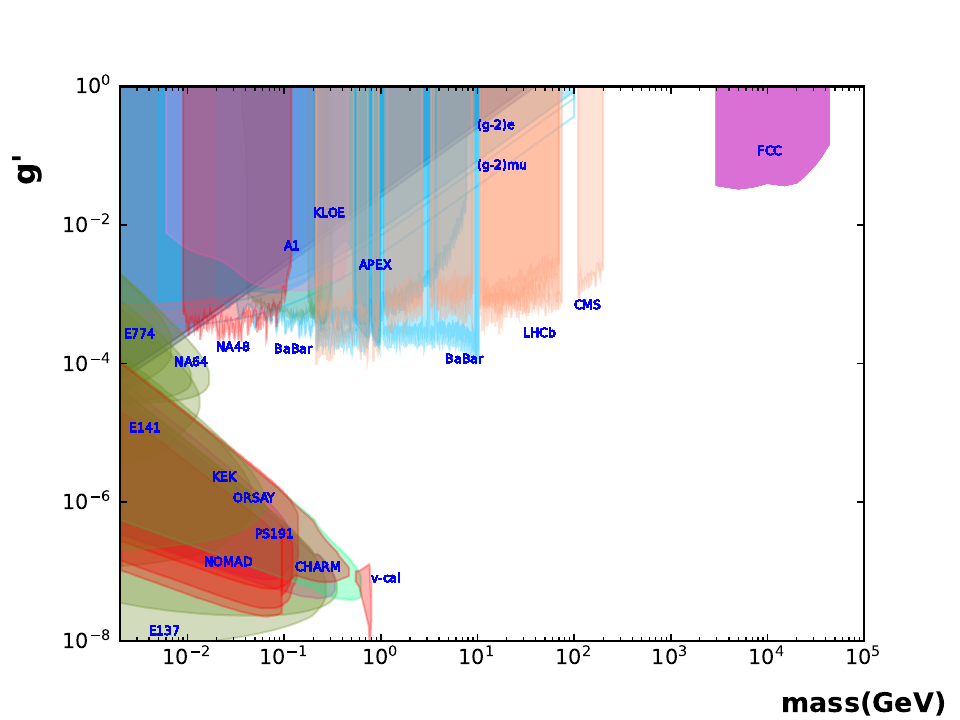}
\caption{Limits of the coupling and mass of $Z'$ with respect to previous experiments and the extended limits of this bounds  determined from the simulation.}
\label{fig3}
\end{figure}

Figure~\ref{fig3} presents the limits on the \(Z'\) coupling and mass, incorporating our new simulation results alongside the existing experimental constraints. The exclusion regions shown in the plot are obtained using \textsf{DARKCAST}~\cite{Ilten:2018crw}, which we employed to reinterpret data from previous dark-photon experiments in the context of the \(B\!-\!L\) model. In particular, the software allows for a model-independent translation of published limits on the kinetic-mixing parameter \(\epsilon\) of a dark photon into corresponding limits on the gauge coupling \(g'\) of a \(B\!-\!L\) boson by rescaling the production and decay rates. This procedure provides a unified framework for comparing dark-photon searches with other \(U(1)_X\) extensions of the Standard Model. Each shaded region represents a parameter space excluded by previous experiments under different physical assumptions. The low-mass domain (\( m_{Z'} \lesssim 10~\text{GeV} \)) is primarily constrained by fixed-target and beam-dump experiments, whereas the intermediate and high-mass ranges are probed by collider and precision measurements. The dark green, red, and brown shaded regions at the lower left correspond to classic beam-dump and neutrino experiments such as E137, E141, PS191, CHARM, and NOMAD, which exclude extremely small couplings (\( g' \lesssim 10^{-7} \)) for sub-GeV masses. Experiments such as APEX, A1, KLOE, NA48, and BaBar probe visible decays at low-mass region, excluding couplings in the range \( 10^{-4} \lesssim g' \lesssim 10^{-2} \). At higher masses (\( m_{Z'} \gtrsim 10~\text{GeV} \)), collider experiments dominate the sensitivity. The light-blue and orange shaded areas represent exclusion regions obtained by LHCb, CMS, and BaBar, which probe the dilepton final states through resonance searches. The CMS limits, in particular, extend up to a few~TeV, excluding couplings above \( g' \sim 10^{-2} \). The magenta region labeled FCC shows the new sensitivity derived in this work using \( 30~\text{ab}^{-1} \) of simulated FCC-hh data. The projected exclusion demonstrates a substantial improvement over existing collider bounds, extending the reachable mass scale up to \( \mathcal{O}(10^{4})~\text{GeV} \) while probing couplings down to \( g' \sim 10^{-2} \). This level of reach reflects the combined effect of the extremely high collision energy and integrated luminosity envisioned for the FCC.

The comparison highlights how future high-energy colliders will complement the low-energy precision and fixed-target programs. Whereas previous experiments have already closed much of the parameter space for light \( Z' \) bosons, the FCC will open an entirely new sensitivity window in the multi-TeV regime, testing the \( B\!-\!L \) hypothesis and other \( U(1)_X \) extensions that predict heavy gauge bosons with electroweak-scale couplings. Importantly, the FCC region shown here bridges the gap between the strongest LEP and LHC exclusions and the currently unexplored parameter space where a \(Z'\) may act as a mediator to the dark sector, with prospective ILC measurements offering complementary indirect sensitivity via high-precision \(e^+e^- \to f\bar{f}\) observables. It is also important to emphasize that the high-mass portion of the FCC exclusion contour extends far beyond the sensitivity of the High-Luminosity LHC and any currently proposed LHC upgrades, underscoring the crucial role of a 100~TeV collider in exploring the vast parameter space of the \(B\!-\!L\) and other \(U(1)\) extensions of the Standard Model.

\section{Summary and Conclusions}
\label{SC}

The \(B\!-\!L\) (Baryon minus Lepton number) extension of the Standard Model offers an elegant framework that naturally incorporates the seesaw mechanism, thereby providing a consistent explanation for the origin of neutrino masses. This model predicts the existence of a new neutral gauge boson, denoted as \(Z'\), whose mass and coupling strength remain experimentally unconstrained over a wide range of parameter space. Establishing limits on these parameters is essential for testing the viability of \(B\!-\!L\) and other \(U(1)_X\) gauge extensions of the Standard Model.

In this work, we investigated the sensitivity of the Future Circular Collider (FCC-hh) to a heavy \(Z'\) boson in the leptonic (\(e^+e^-,\, \mu^+\mu^-\)) final states. The analysis focused exclusively on the dominant Standard Model Drell--Yan background, which rapidly decreases with increasing invariant mass, providing an almost background-free region in the multi-TeV regime. Using simulated FCC data corresponding to an integrated luminosity of \(30~\text{ab}^{-1}\), we derived 95\% CL upper limits on the production cross section \(\sigma(pp \to Z') \times \mathrm{BR}(Z' \to \ell^+\ell^-)\) for various coupling constants. For the largest coupling considered (\(g' = 0.3\)), the analysis excludes \(Z'\) masses up to approximately \(45~\text{TeV}\), demonstrating the exceptional reach of a 100~TeV collider in probing physics beyond the Standard Model.

To place these results in a broader experimental context, we employed the \textsf{DARKCAST} framework to reinterpret existing dark-photon limits in terms of the \(B\!-\!L\) gauge coupling. By recalibrating the constraints from previous beam-dump, fixed-target, and collider experiments, we established a unified comparison of the existing exclusion regions with our new FCC projection. The results show that the FCC significantly extends the accessible parameter space, bridging the gap between the strongest current limits from LEP and the LHC, and entering a previously unexplored regime where a heavy \(Z'\) could still mediate interactions between the Standard Model and the dark sector.

In summary, this study demonstrates that the FCC-hh, with its unprecedented center-of-mass energy and luminosity, provides a powerful environment to explore heavy neutral gauge bosons predicted by \(B\!-\!L\) and related \(U(1)_X\) extensions. The derived limits represent a substantial improvement over existing bounds and establish a strong case for future experimental efforts to test the presence of new gauge interactions at the multi-TeV scale.

\section*{Acknowledgments}
We gratefully acknowledge the Office of the Vice Chancellor for Research and Extension – Department of Research (OVCRE-DR) of Mindanao State University – Iligan Institute of Technology for providing research funding. We also extend our sincere thanks to the Department of Science and Technology – Accelerated Science and Technology Human Resource Development Program (DOST-ASTHRDP) for the scholarship support. Additionally, we would like to thank Jennifer Kile for her insightful discussions on the $CL_S$ method and for recommending valuable references.

\normalsize
\bibliography{references}

@article{ALEPH:2013dgf,
  author        = "Schael, S. and others",
  collaboration = "ALEPH and DELPHI and L3 and OPAL and LEP Electroweak",
  title         = "{Electroweak Measurements in Electron-Positron Collisions at W-Boson-Pair Energies at LEP}",
  journal       = "Phys. Rept.",
  year          = "2013",
  doi           = "10.1016/j.physrep.2013.07.004"
}

@article{Cacciapaglia:2006,
  author        = "Cacciapaglia, G. and Cs{\'a}ki, C. and Marandella, G. and Strumia, A.",
  title         = "{The minimal set of electroweak precision parameters}",
  journal       = "Phys. Rev. D",
  volume        = "74",
  number        = "3",
  pages         = "033011",
  year          = "2006",
  doi           = "10.1103/PhysRevD.74.033011"
}

@article{Das:2021esm,
  author        = "Das, Arindam and Dev, P. S. Bhupal and Hosotani, Yutaka and Mandal, Sanjoy",
  title         = "{Probing the minimal $U(1)_X$ model at future electron-positron colliders via fermion pair-production channels}",
  journal       = "Phys. Rev. D",
  volume        = "105",
  number        = "11",
  pages         = "115030",
  year          = "2022",
  doi           = "10.1103/PhysRevD.105.115030",
  eprint        = "2104.10902",
  archivePrefix = "arXiv",
  primaryClass  = "hep-ph"
}

@article{A1:2011yso,
  author        = "Merkel, H. and others",
  collaboration = "A1",
  title         = "{Search for Light Gauge Bosons of the Dark Sector at the Mainz Microtron}",
  journal       = "Phys. Rev. Lett.",
  volume        = "106",
  pages         = "251802",
  year          = "2011",
  doi           = "10.1103/PhysRevLett.106.251802",
  eprint        = "1101.4091",
  archivePrefix = "arXiv",
  primaryClass  = "nucl-ex"
}

@article{NA64:2019auh,
  author        = "Banerjee, D. and others",
  collaboration = "NA64",
  title         = "{Improved limits on a hypothetical $X(16.7)$ boson and a dark photon decaying into $e^+e^-$ pairs}",
  journal       = "Phys. Rev. D",
  volume        = "101",
  number        = "7",
  pages         = "071101",
  year          = "2020",
  doi           = "10.1103/PhysRevD.101.071101",
  eprint        = "1912.11389",
  archivePrefix = "arXiv",
  primaryClass  = "hep-ex",
  reportNumber  = "CERN-EP-2019-284"
}

@article{NA64:2017vtt,
  author        = "Banerjee, D. and others",
  collaboration = "NA64",
  title         = "{Search for vector mediator of Dark Matter production in invisible decay mode}",
  journal       = "Phys. Rev. D",
  volume        = "97",
  number        = "7",
  pages         = "072002",
  year          = "2018",
  doi           = "10.1103/PhysRevD.97.072002",
  eprint        = "1710.00971",
  archivePrefix = "arXiv",
  primaryClass  = "hep-ex"
}

@article{KLOE-2:2016ydq,
  author        = "Anastasi, A. and others",
  collaboration = "KLOE-2",
  title         = "{Limit on the production of a new vector boson in $e^+e^- \to U\gamma$, $U \to \pi^+\pi^-$ with the KLOE experiment}",
  journal       = "Phys. Lett. B",
  volume        = "757",
  pages         = "356--361",
  year          = "2016",
  doi           = "10.1016/j.physletb.2016.04.019",
  eprint        = "1603.06086",
  archivePrefix = "arXiv",
  primaryClass  = "hep-ex"
}

@article{APEX:2011dww,
  author        = "Abrahamyan, S. and others",
  collaboration = "APEX",
  title         = "{Search for a New Gauge Boson in Electron-Nucleus Fixed-Target Scattering by the APEX Experiment}",
  journal       = "Phys. Rev. Lett.",
  volume        = "107",
  pages         = "191804",
  year          = "2011",
  doi           = "10.1103/PhysRevLett.107.191804",
  eprint        = "1108.2750",
  archivePrefix = "arXiv",
  primaryClass  = "hep-ex",
  reportNumber  = "JLAB-PHY-11-1406, SLAC-PUB-14491"
}

@article{Asai:2023mzl,
  author        = "Asai, Kento and Das, Arindam and Li, Jinmian and Nomura, Takaaki and Seto, Osamu",
  title         = "{Probing for chiral $Z'$ gauge boson through scattering measurement experiments}",
  journal       = "Phys. Rev. D",
  volume        = "109",
  number        = "7",
  pages         = "075026",
  year          = "2024",
  doi           = "10.1103/PhysRevD.109.075026",
  eprint        = "2307.09737",
  archivePrefix = "arXiv",
  primaryClass  = "hep-ph",
  reportNumber  = "EPHOU-23-013"
}

@article{Essig:2010xa,
  author        = "Essig, Rouven and Schuster, Philip and Toro, Natalia and Wojtsekhowski, Bogdan",
  title         = "{An Electron Fixed Target Experiment to Search for a New Vector Boson $A'$ Decaying to $e^+e^-$}",
  journal       = "JHEP",
  volume        = "02",
  pages         = "009",
  year          = "2011",
  doi           = "10.1007/JHEP02(2011)009",
  eprint        = "1001.2557",
  archivePrefix = "arXiv",
  primaryClass  = "hep-ph",
  reportNumber  = "SLAC-PUB-13882, SU-ITP-10-01"
}

@article{Freytsis:2009bh,
  author        = "Freytsis, Marat and Ovanesyan, Grigory and Thaler, Jesse",
  title         = "{Dark Force Detection in Low Energy $e^- p$ Collisions}",
  journal       = "JHEP",
  volume        = "01",
  pages         = "111",
  year          = "2010",
  doi           = "10.1007/JHEP01(2010)111",
  eprint        = "0909.2862",
  archivePrefix = "arXiv",
  primaryClass  = "hep-ph"
}

@article{BaBar:2014zli,
  author        = "Lees, J. P. and others",
  collaboration = "BaBar",
  title         = "{Search for a Dark Photon in $e^+e^-$ Collisions at BaBar}",
  journal       = "Phys. Rev. Lett.",
  volume        = "113",
  number        = "20",
  pages         = "201801",
  year          = "2014",
  doi           = "10.1103/PhysRevLett.113.201801",
  eprint        = "1406.2980",
  archivePrefix = "arXiv",
  primaryClass  = "hep-ex",
  reportNumber  = "BABAR-PUB-14-002, SLAC-PUB-15979"
}

@article{BESIII:2017fwv,
  author        = "Ablikim, M. and others",
  collaboration = "BESIII",
  title         = "{Dark Photon Search in the Mass Range Between 1.5 and 3.4 GeV/$c^2$}",
  journal       = "Phys. Lett. B",
  volume        = "774",
  pages         = "252--257",
  year          = "2017",
  doi           = "10.1016/j.physletb.2017.09.067",
  eprint        = "1705.04265",
  archivePrefix = "arXiv",
  primaryClass  = "hep-ex"
}

@article{LHCb:2017trq,
  author        = "Aaij, R. and others",
  collaboration = "LHCb",
  title         = "{Search for Dark Photons Produced in 13 TeV $pp$ Collisions}",
  journal       = "Phys. Rev. Lett.",
  volume        = "120",
  number        = "6",
  pages         = "061801",
  year          = "2018",
  doi           = "10.1103/PhysRevLett.120.061801",
  eprint        = "1710.02867",
  archivePrefix = "arXiv",
  primaryClass  = "hep-ex",
  reportNumber  = "LHCB-PAPER-2017-038, CERN-EP-2017-248"
}

@article{CMS:2020ulv,
  author        = "Sirunyan, A. M. and others",
  collaboration = "CMS",
  title         = "{Search for dark matter produced in association with a leptonically decaying $Z$ boson in proton-proton collisions at $\sqrt{s} = 13$ TeV}",
  journal       = "Eur. Phys. J. C",
  volume        = "81",
  number        = "1",
  pages         = "13",
  year          = "2021",
  doi           = "10.1140/epjc/s10052-020-08739-5",
  eprint        = "2008.04735",
  archivePrefix = "arXiv",
  primaryClass  = "hep-ex",
  reportNumber  = "CMS-EXO-19-003, CERN-EP-2020-136",
  note          = "[Erratum: Eur. Phys. J. C 81, 333 (2021)]"
}

@article{CMS:2024zqs,
  author        = "Hayrapetyan, Aram and others",
  collaboration = "CMS",
  title         = "{Dark sector searches with the CMS experiment}",
  journal       = "Phys. Rept.",
  volume        = "1115",
  pages         = "448--569",
  year          = "2025",
  doi           = "10.1016/j.physrep.2024.09.013",
  eprint        = "2405.13778",
  archivePrefix = "arXiv",
  primaryClass  = "hep-ex",
  reportNumber  = "CMS-EXO-23-005, CERN-EP-2024-106"
}

@article{KA:2023dyz,
  author        = "K. A., ShivaSankar and Das, Arindam and Lambiase, Gaetano and Nomura, Takaaki and Orikasa, Yuta",
  title         = "{Probing chiral and flavored $Z'$ from cosmic bursts through neutrino interactions}",
  journal       = "Eur. Phys. J. C",
  volume        = "84",
  number        = "11",
  pages         = "1224",
  year          = "2024",
  doi           = "10.1140/epjc/s10052-024-13530-x",
  eprint        = "2308.14483",
  archivePrefix = "arXiv",
  primaryClass  = "hep-ph"
}

@article{Barman:2024lxy,
  author        = "Barman, Basabendu and Das, Arindam and Mandal, Sanjoy",
  title         = "{Dark matter-electron scattering and freeze-in scenarios in the light of $Z'$ mediation}",
  journal       = "Phys. Rev. D",
  volume        = "110",
  number        = "5",
  pages         = "055029",
  year          = "2024",
  doi           = "10.1103/PhysRevD.110.055029",
  eprint        = "2407.00969",
  archivePrefix = "arXiv",
  primaryClass  = "hep-ph"
}

@article{Barman:2025bir,
  author        = "Barman, Basabendu and Das, Arindam and Das, Suruj Jyoti and Merchand, Marco",
  title         = "{Hunting for heavy $Z'$ with IceCube neutrinos and gravitational waves}",
  journal       = "Phys. Rev. D",
  volume        = "112",
  number        = "3",
  pages         = "035035",
  year          = "2025",
  eprint        = "2502.13217",
  archivePrefix = "arXiv",
  primaryClass  = "hep-ph"
}

@article{Barman:2025hoz,
  author        = "Barman, Basabendu and Das, Arindam and Sarmah, Prantik",
  title         = "{What KM3-230213A events may tell us about the neutrino mass and dark matter}",
  year          = "2025",
  eprint        = "2504.01447",
  archivePrefix = "arXiv",
  primaryClass  = "hep-ph"
}

@article{Babich:2010zz,
  author        = "Babich, A. A. and Pankov, A. A. and Tsytrinov, A. V. and Karpenko, N. V.",
  title         = "{Searches for new neutral gauge $Z'$ bosons at the $e^+ e^-$ International Linear Collider and their identification}",
  journal       = "Phys. Atom. Nucl.",
  volume        = "73",
  pages         = "773--784",
  year          = "2010",
  doi           = "10.1134/S1063778810050054"
}

@article{Pankov:2017dkv,
  author        = "Pankov, A. A. and Tsytrinov, A. V.",
  title         = "{Model identification of new heavy $Z'$ bosons at ILC with polarized beams}",
  journal       = "J. Phys. Conf. Ser.",
  volume        = "938",
  number        = "1",
  pages         = "012059",
  year          = "2017",
  doi           = "10.1088/1742-6596/938/1/012059"
}

@article{Helsens:2019bfw,
  author        = "Helsens, C. and Jamin, D. and Mangano, M. L. and Rizzo, T. G. and Selvaggi, M.",
  title         = "{Heavy resonances at energy-frontier hadron colliders}",
  journal       = "Eur. Phys. J. C",
  volume        = "79",
  pages         = "569",
  year          = "2019",
  doi           = "10.1140/epjc/s10052-019-7062-3",
  eprint        = "1902.11217",
  archivePrefix = "arXiv",
  primaryClass  = "hep-ph",
  reportNumber  = "CERN-TH-2019-020, SLAC-PUB-17408"
}

@article{FCC:2018byv,
    author = "Abada, A. and others",
    collaboration = "FCC",
    title = "{FCC Physics Opportunities}: {Future Circular Collider Conceptual Design Report Volume 1}",
    reportNumber = "CERN-ACC-2018-0056",
    doi = "10.1140/epjc/s10052-019-6904-3",
    journal = "Eur. Phys. J. C",
    volume = "79",
    number = "6",
    pages = "474",
    year = "2019"
}

@article{FCC:2018vvp,
    author = "Abada, A. and others",
    collaboration = "FCC",
    title = "{FCC-hh: The Hadron Collider}: {Future Circular Collider Conceptual Design Report Volume 3}",
    reportNumber = "CERN-ACC-2018-0058",
    doi = "10.1140/epjst/e2019-900087-0",
    journal = "Eur. Phys. J. ST",
    volume = "228",
    number = "4",
    pages = "755--1107",
    year = "2019"
}

@article{FCC:2018evy,
    author = "Abada, A. and others",
    collaboration = "FCC",
    title = "{FCC-ee: The Lepton Collider}: {Future Circular Collider Conceptual Design Report Volume 2}",
    reportNumber = "CERN-ACC-2018-0057",
    doi = "10.1140/epjst/e2019-900045-4",
    journal = "Eur. Phys. J. ST",
    volume = "228",
    number = "2",
    pages = "261--623",
    year = "2019"
}

@article{Apollinari:2015wtw,
    author = {Apollinari, G. and Br\"uning, O. and Nakamoto, T. and Rossi, Lucio},
    editor = {Apollinari, G and B\'ejar Alonso, I and Br\"uning, O and Lamont, M and Rossi, L},
    title = "{High Luminosity Large Hadron Collider HL-LHC}",
    eprint = "1705.08830",
    archivePrefix = "arXiv",
    primaryClass = "physics.acc-ph",
    reportNumber = "FERMILAB-PUB-15-699-TD",
    doi = "10.5170/CERN-2015-005.1",
    journal = "CERN Yellow Rep.",
    number = "5",
    pages = "1--19",
    year = "2015"
}

@article{LHeC:2020van,
    author = "Agostini, P. and others",
    collaboration = "LHeC, FCC-he Study Group",
    title = "{The Large Hadron\textendash{}Electron Collider at the HL-LHC}",
    eprint = "2007.14491",
    archivePrefix = "arXiv",
    primaryClass = "hep-ex",
    reportNumber = "CERN-ACC-Note-2020-0002, JLAB-ACP-20-3180",
    doi = "10.1088/1361-6471/abf3ba",
    journal = "J. Phys. G",
    volume = "48",
    number = "11",
    pages = "110501",
    year = "2021"
}

@article{FCC:2018bvk,
    author = "Abada, A. and others",
    collaboration = "FCC",
    title = "{HE-LHC: The High-Energy Large Hadron Collider}: {Future Circular Collider Conceptual Design Report Volume 4}",
    reportNumber = "CERN-ACC-2018-0059",
    doi = "10.1140/epjst/e2019-900088-6",
    journal = "Eur. Phys. J. ST",
    volume = "228",
    number = "5",
    pages = "1109--1382",
    year = "2019"
}

@article{CEPC-SPPCStudyGroup:2015csa,
    author = "Ahmad, Muhammd and others",
    title = "{CEPC-SPPC Preliminary Conceptual Design Report. 1. Physics and Detector}",
    reportNumber = "IHEP-CEPC-DR-2015-01, IHEP-TH-2015-01, IHEP-EP-2015-01",
    month = "3",
    year = "2015"
}

@article{CEPCStudyGroup:2023quu,
    author = "Abdallah, Waleed and others",
    collaboration = "CEPC Study Group",
    title = "{CEPC Technical Design Report -- Accelerator (v2)}",
    eprint = "2312.14363",
    archivePrefix = "arXiv",
    primaryClass = "physics.acc-ph",
    reportNumber = "IHEP-CEPC-DR-2023-01, IHEP-AC-2023-01",
    month = "12",
    year = "2023"
}

@article{Armesto:2014iaa,
    author = "Armesto, Nestor and Dainese, Andrea and d'Enterria, David and Masciocchi, Silvia and Roland, Christof and Salgado, Carlos and van Leeuwen, Marco and Wiedemann, Urs",
    editor = "Braun-Munzinger, Peter and Friman, Bengt and Stachel, Johanna",
    title = "{Heavy-ion physics studies for the Future Circular Collider}",
    eprint = "1407.7649",
    archivePrefix = "arXiv",
    primaryClass = "nucl-ex",
    doi = "10.1016/j.nuclphysa.2014.09.067",
    journal = "Nucl. Phys. A",
    volume = "931",
    pages = "1163--1168",
    year = "2014"
}

@article{Basso:2010pe,
    author = "Basso, Lorenzo and Belyaev, Alexander and Moretti, Stefano and Pruna, Giovanni Marco and Shepherd-Themistocleous, Claire H.",
    title = "{$Z'$ discovery potential at the LHC in the minimal $B-L$ extension of the Standard Model}",
    eprint = "1002.3586",
    archivePrefix = "arXiv",
    primaryClass = "hep-ph",
    reportNumber = "SHEP-09-19, DFTT-58-2009",
    doi = "10.1140/epjc/s10052-011-1613-6",
    journal = "Eur. Phys. J. C",
    volume = "71",
    pages = "1613",
    year = "2011"
}

@article{Basso:2008iv,
    author = "Basso, Lorenzo and Belyaev, Alexander and Moretti, Stefano and Shepherd-Themistocleous, Claire H.",
    title = "{Phenomenology of the minimal B-L extension of the Standard model: Z' and neutrinos}",
    eprint = "0812.4313",
    archivePrefix = "arXiv",
    primaryClass = "hep-ph",
    reportNumber = "SHEP-08-13",
    doi = "10.1103/PhysRevD.80.055030",
    journal = "Phys. Rev. D",
    volume = "80",
    pages = "055030",
    year = "2009"
}

@article{Basso:2010jm,
    author = "Basso, Lorenzo and Moretti, Stefano and Pruna, Giovanni Marco",
    title = "{A Renormalisation Group Equation Study of the Scalar Sector of the Minimal B-L Extension of the Standard Model}",
    eprint = "1004.3039",
    archivePrefix = "arXiv",
    primaryClass = "hep-ph",
    reportNumber = "SHEP-10-16",
    doi = "10.1103/PhysRevD.82.055018",
    journal = "Phys. Rev. D",
    volume = "82",
    pages = "055018",
    year = "2010"
}

@article{Alwall:2014hca,
    author = "Alwall, J. and Frederix, R. and Frixione, S. and Hirschi, V. and Maltoni, F. and Mattelaer, O. and Shao, H. -S. and Stelzer, T. and Torrielli, P. and Zaro, M.",
    title = "{The automated computation of tree-level and next-to-leading order differential cross sections, and their matching to parton shower simulations}",
    eprint = "1405.0301",
    archivePrefix = "arXiv",
    primaryClass = "hep-ph",
    reportNumber = "CERN-PH-TH-2014-064, CP3-14-18, LPN14-066, MCNET-14-09, ZU-TH-14-14",
    doi = "10.1007/JHEP07(2014)079",
    journal = "JHEP",
    volume = "07",
    pages = "079",
    year = "2014"
}

@article{NNPDF:2014otw,
    author = "Ball, Richard D. and others",
    collaboration = "NNPDF",
    title = "{Parton distributions for the LHC Run II}",
    eprint = "1410.8849",
    archivePrefix = "arXiv",
    primaryClass = "hep-ph",
    reportNumber = "EDINBURGH-2014-15, IFUM-1034-FT, CERN-PH-TH-2013-253, OUTP-14-11P, CAVENDISH-HEP-14-11",
    doi = "10.1007/JHEP04(2015)040",
    journal = "JHEP",
    volume = "04",
    pages = "040",
    year = "2015"
}

@article{deFavereau:2013fsa,
    author = "De Favereau, J. and Delaere, C. and Demin, P. and Giammanco, A. and Lema\^\i{}tre, V. and Mertens, A. and Selvaggi, M.",
    collaboration = "DELPHES 3",
    title = "{DELPHES 3, A modular framework for fast simulation of a generic collider experiment}",
    eprint = "1307.6346",
    archivePrefix = "arXiv",
    primaryClass = "hep-ex",
    doi = "10.1007/JHEP02(2014)057",
    journal = "JHEP",
    volume = "02",
    pages = "057",
    year = "2014"
}

@article{Cowan:2010js,
    author = "Cowan, Glen and Cranmer, Kyle and Gross, Eilam and Vitells, Ofer",
    title = "{Asymptotic formulae for likelihood-based tests of new physics}",
    eprint = "1007.1727",
    archivePrefix = "arXiv",
    primaryClass = "physics.data-an",
    doi = "10.1140/epjc/s10052-011-1554-0",
    journal = "Eur. Phys. J. C",
    volume = "71",
    pages = "1554",
    year = "2011",
    note = "[Erratum: Eur.Phys.J.C 73, 2501 (2013)]"
}

@article{Moneta:2010pm,
    author = "Moneta, Lorenzo and Belasco, Kevin and Cranmer, Kyle S. and Kreiss, S. and Lazzaro, Alfio and Piparo, Danilo and Schott, Gregory and Verkerke, Wouter and Wolf, Matthias",
    editor = "Speer, T. and Boudjema, F. and Lauret, Jerome and Naumann, Axel and Teodorescu, L. and Uwer, P.",
    title = "{The RooStats Project}",
    eprint = "1009.1003",
    archivePrefix = "arXiv",
    primaryClass = "physics.data-an",
    doi = "10.22323/1.093.0057",
    journal = "PoS",
    volume = "ACAT2010",
    pages = "057",
    year = "2010"
}

@article{Ilten:2018crw,
    author = "Ilten, Philip and Soreq, Yotam and Williams, Mike and Xue, Wei",
    title = "{Serendipity in dark photon searches}",
    eprint = "1801.04847",
    archivePrefix = "arXiv",
    primaryClass = "hep-ph",
    reportNumber = "MIT-CTP/4976, CERN-TH-2017-282, MIT-CTP-4976",
    doi = "10.1007/JHEP06(2018)004",
    journal = "JHEP",
    volume = "06",
    pages = "004",
    year = "2018"
}

@article{Abbas:2007ag,
    author = "Abbas, M. and Khalil, S.",
    title = "{Neutrino masses, mixing and leptogenesis in TeV scale $B$ - L extension of the standard model}",
    eprint = "0707.0841",
    archivePrefix = "arXiv",
    primaryClass = "hep-ph",
    doi = "10.1088/1126-6708/2008/04/056",
    journal = "JHEP",
    volume = "04",
    pages = "056",
    year = "2008"
}

@article{Iso:2010mv,
    author = "Iso, Satoshi and Okada, Nobuchika and Orikasa, Yuta",
    title = "{Resonant Leptogenesis in the Minimal B-L Extended Standard Model at TeV}",
    eprint = "1011.4769",
    archivePrefix = "arXiv",
    primaryClass = "hep-ph",
    reportNumber = "KEK-TH-1422",
    doi = "10.1103/PhysRevD.83.093011",
    journal = "Phys. Rev. D",
    volume = "83",
    pages = "093011",
    year = "2011"
}

@article{Hernandez:2025spl,
    author = "Hern{\'a}ndez, Pilar",
    editor = "Elsing, Markus and Huss, Alexander Yohei",
    title = "{Neutrino physics}",
    doi = "10.23730/CYRSP-2025-001.139",
    journal = "CERN Yellow Rep. School Proc.",
    volume = "1",
    pages = "139",
    year = "2025"
}

@article{Hindmarsh:2020hop,
    author = {Hindmarsh, Mark B. and L{\"u}ben, Marvin and Lumma, Johannes and Pauly, Martin},
    title = "{Phase transitions in the early universe}",
    eprint = "2008.09136",
    archivePrefix = "arXiv",
    primaryClass = "astro-ph.CO",
    reportNumber = "MPP-2020-163, HIP-2020-27/TH",
    doi = "10.21468/SciPostPhysLectNotes.24",
    journal = "SciPost Phys. Lect. Notes",
    volume = "24",
    pages = "1",
    year = "2021"
}

@phdthesis{Seuthe:2022vea,
    author = "Seuthe, Alex",
    title = "{Test of lepton flavour universality with rare beauty-quark decays at the LHCb experiment}",
    doi = "10.17877/DE290R-22834",
    school = {Technischen Universit{\"a}t Dortmund, Tech. U., Dortmund (main)},
    year = "2022"
}

@article{Bodas:2021fsy,
    author = "Bodas, Arushi and Coy, Rupert and King, Simon J. D.",
    title = "{Solving the electron and muon $g-2$ anomalies in $Z'$ models}",
    eprint = "2102.07781",
    archivePrefix = "arXiv",
    primaryClass = "hep-ph",
    reportNumber = "ULB-TH/21-01",
    doi = "10.1140/epjc/s10052-021-09850-x",
    journal = "Eur. Phys. J. C",
    volume = "81",
    number = "12",
    pages = "1065",
    year = "2021"
}


\end{document}